# Technological integration and hyper-connectivity: Tools for promoting extreme human lifespans

Marios Kyriazis

**Abstract**

Artificial, neurobiological, and social networks are three distinct complex adaptive systems (CAS), each containing discrete processing units (nodes, neurons, and humans respectively). Despite the apparent differences, these three networks are bound by common underlying principles which describe the behaviour of the system in terms of the connections of its components, and its emergent properties. The longevity (long-term retention and functionality) of the components of each of these systems is also defined by common principles. Here, I will examine some properties of the longevity and function of the components of artificial and neurobiological systems, and generalise these to the longevity and function of the components of social CAS. In other words, I will show that principles governing the long-term functionality of computer nodes and of neurons, may be extrapolated to the study of the long-term functionality of humans(or more precisely,of the **noemes**, an abstract combination of 'existence' and 'digital fame'). The study of these phenomena can provide useful insights regarding practical ways that can be used in order to maximize human longevity.The basic law governing these behaviours is the 'Law of Requisite Usefulness', which states that the length of retention of an agent within a CAS is proportional to the agent's contribution to the overall adaptability of the system.

**Introduction**

Evolution is driven by non-equilibrium processes which collectively increase both the entropy and the information content of the species (1). Evolution (i.e. increased complexity and organisation useful withina specified environment) can occur both with and without environmental or sexual selection (1). Here, an attempt will be made to study the common evolutionary mechanisms found in some self-organisingcomplex adaptive systems (CAS), namely artificial networks, the human brain, and the Global Brain. The Global Brain (GB) is the worldwide network formed by the combined distributed intelligence of people, information and communication technologies that connect them into a self-organised system (2). It is a complex adaptive system displaying properties that emerge from the network interactions between its individual components. Information within the GB propagates according to the same basic rules as those encountered both in computer networks and in the human brain (3).

Complexity theory shows that there are deep underlying similarities in concepts and processes, between completely different fields (4). For instance, it helps us study the similar priniples that exist between computers (artificial networks), neurons (biological networks) and human society (such as the GB and it individual constituents the 'noemes', a term which will be described below).These relationships are transferable across different levels of hierarchy, making it relatively easy to speculate and/or predict behaviours of higher systems. In this paper I adopt a systems biological emphasis, with an ultimate aim to explore certain aspects of the process of human ageing. Systems biology is an increasingly important biological science, which may provide significant insights into the process of ageing. It explains the behaviour of a biological entity using a 'connection-oriented' approach, rather than a reductionist 'component-oriented' one. In other words, it examines how different components are connected and regulated rather than studying the properties of single individual agents. Systems biology studies the emergent properties of the components of a living system, properties which cannot simply be reduced to the sum of its individual parts alone. Using such an approach it may be appropriate to discuss how individual units behave within an organised system, in an attempt to examine the phenomenon of ageing from an alternative standpoint. Specifically, I examine how three distinct agents (i.e. autonomous, goal-directed, information-processing actors who act on their environment) behave within:

1. An artificial network (the agents here are computer nodes i.e. devices connected to a computer network).

2. The brain (the agents are the neurons), and

3. The Global Brain (the agents are the 'noemes', i.e. humans integrated with technology).

All three agents have common characteristics, namely:

- Input (information collection)
- Processing (interpretation of data, and problem-solving)
- Output (efficient and relevant action)
- Feedback (capacity to learn and adapt)

The behaviour of these autonomous agents contributes to the overall fitness of the system, and modulates the adaptability of the system, ultimately defining its survival. It is worth noting that my definition of 'fitness' in this paper is not the traditional Darwinian one. Instead fitness is defined as 'resilience during change' or 'an adaptation to an environment whose complexity co-evolves with the complexity of the system'. A system which adapts and improves its performance as its environment changes, is a 'fit' system and it can survive longer (Figure 1).

\*\*\*\*\*\*\*\*\*\*\*\*\*\*\*\*\*\*\*\*\*\*\*\*\*\*\*\*\*\*\*\*\*\*\*\*\*\*\*\*\*\*\*\*\*\*\*\*\*\*\*\*\*\*\*\*\*\*\*\*\*\*\*\*\*\*\*\*\*\*\*\*\*\*\*\*\*\*\*\*\*\*

FIGURE1

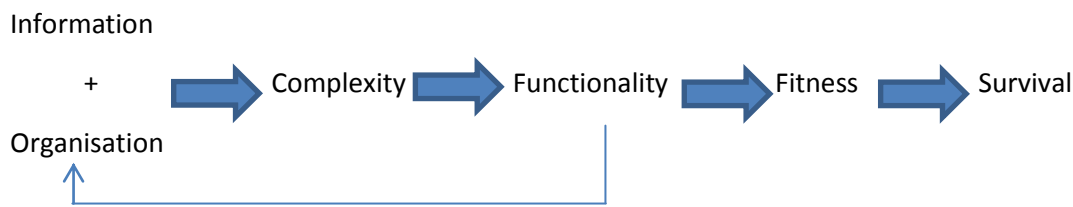

Figure 1.(from Kyriazis M. (5) Information (plus organisation) increases complexity and this increases functionality. This improves fitness and thus survival. (Functionality also increases organisation). Any increase in internal fitness requires the formation of new links and the strengthening of the

interconnectedness between its nodes i.e. increased complexity, thus increased fitness and increased survival

*******************************************************************************

It is possible to study some from universal aspects of the evolutionary process by using examples from the computational sciences (6). In an information-rich environment which is subjected to finite allocation of resources, nodes(computer programmes, i.e. 'digital organisms') can survive (retained via self-replication), evolve and adapt to their environment, and incorporate progressively more information into their systems. By contrast, nodes found in information-poor landscapes shrink with time and, in order to survive, resort to replication only (i.e.there is no growth or development, and no adaptation to their environment).Successful survival in a system (i.e. a stable entropy state) can be achieved by maintaining an extreme level of information. Here, the definition of information is 'a meaningful set of data or patterns which influence the formation or transformation of other data or patterns, in order to reduce uncertainty and help achieve a goal' (7). Therefore, information ultimately leads to survival (figure 1).This is highlighted further by considering a universal cybernetic concept. Ashby's Law of Requisite Variety (8) posits that organisms must increase their own complexity (i.e. their information content needed for fitness) when these organisms are found inless predictable (i.e. more complex) environments. As a result, considering that our technological environment is becomingincreasingly more complex, we too need to increase our information content andfunctional complexity in order to be able to survive within this specific environment (5).

In network theory, low activity nodes interact preferentially with high activity ones anywhere in the network. This principle is also utilised by individuals for optimising social networks. Due to selection pressures in growing groups, individuals aim to build connections with other well-accepted individuals, and to abandon connections with less popular group members.  Appropriate perturbations (such as an increase of meaningful information) can cause a system to move away from equilibrium (and thus from stagnation) to a state of evolutionary exploration, facilitating new and diverse adaptations. However, over time, both the components of a system and their interactions move again towards local optima (stagnant landscapes) where network connectivity decreases. It is necessary to continue the perturbation(such as high-level information sharing) in order to lead to more new adaptations, and thus continuing survival (9). Therefore, it is plausible to suggest that, in order for fitness to be maintained, we need to continually perturb the system, by actively receiving, processing and transmitting information.

The cybernetic concept of 'Selective Reinforcement' provides additional theoretical support for the above suggestion(10).  The concepts holds that an appropriate agent (or action) is selected and retained if its content of information is of a sufficient and appropriate magnitude.Research has shown(11) that an online avatar can live longer (through user retention) within a virtual society, if this avatar is well connected within that particular society.  What seems to be of importance in the retention of an avatar is both the number and the strength (diversity) of its connections i.e. their 'integration' with the society they live in.User disengagement and short retention of users has been correlated with low quality content (12).The ability to earn online reputation and intrinsic motivation for participating content are also factors that affect retention (13). Therefore, it is not just the mere number of connections (the structural complexity) but also the 'richness' of those connections (the functional complexity) that matter. This is a universal theme also encountered in the retention (survival) of neurons in the brain (14). The concept can be extrapolated to humans who act as autonomous agents within a technological society. At this stage, it is useful to consider the notion of the noeme.

**The noeme**

The noeme *[(plural noemes), pronounced noːïm, from the Greek νόημα which means 'that which emanates from the Nous (brain/mind)']* is the intellectual 'presence' of an individual within the Global Brain (GB). A noeme is an active evolutionary replicator (a similar concept as that of genes and memes). It can be envisaged as a 'networked-self', globally connected to other noemes through digital communications technology, a meaningful synergy and coexistence of humans, their social interactions and artificial agents (15). Just as neurons are the individual agents of the brain, noemes are the individual agents of the GB. The equivalent of the neuronal synapse in the GB is the hyperlink. The noeme ($N$t) is a dynamic, time-dependent amalgam of a person ($P$), the total number of webpages ($W$) that person is connected to, and the associated links ($L$) i.e. **connections** as a function of time (t), which give rise to an entity with emergent characteristics. This entity is described by:

$$N(t) = P.W^n.S + L(t).U \quad \{1\}$$

Here, $S$ is the 'strength' of presence denoted by

$$S = c + ec + r \quad \{2\}$$

    where c is the consistency of the online username

    ec is the expressive content of webpages, and

    r is the online credibility,

and $U$ is the 'unity' of the connections over time, denoted by:

$$U = fc.m.r(t) \quad \{3\}$$

    where

    fc is the frequency of connection

    m is the meaningfulness of connection, and

    r(t) is the repeatability of the connections over time

The noeme is structurally coupled with its medium, i.e. the computer/internet. It continuously generates its own organisation and specifies its operation and content. As a self-organising system it adjusts to external influences and reinvents itself in order to adapt to its environment i.e. it reproduces (self-replicates) horizontally in a process that can be termed '*noemic reproduction*'. This digital intellectual manifestation of a person, if successful, will lead to others copying it, thus noemes are replicating. It is copied (replicates) by imitation and technologically-assisted transfer, for example, via the internet. A noeme is not just a collection of single ideas or solitary intellectual achievements. It is the total sum of all individual cognitive efforts and active information-sharing accomplishments of a person, the intellectual standing of a person within the GB. As complex adaptive systems themselves, noemes are fluid (not static) and continue to evolve and adapt. They can co-evolve in association with genes and memes – a symbiotic relationship. Genes and memes can enhance the presence of a noeme through, for instance, producing a genetically robust brain that has strong intellect and, through memes, consolidating its impact within society. It is possible that there exists a co-evolution of genes, memes and noemes that will result in a functional GB that, in turn, enhances all three. The GB is developing not for the sole benefit of humans but also for a better replication of memes, and also genes (through better health perhaps). Noemes that do not contribute to the fitness of the GB are eventually eliminated. Those who enhance the fitness of the GB are retained and improved. All three can evolve in opposition. For example, a noeme which is not well integrated can result in its elimination (and thus the elimination of its human host) through early ageing anddeath, thus the end of the gene/meme that defined it. Or, a noeme defined by beneficial memes/genes can integrate well in the GB, evolve and transmit their genes/memes to others. Noemes can mutate and result in better, worse or neutral ones. The behaviour of noemes helps us define ourselves in a way that strengthens our rational presence in the world. By trying to enhance our information-sharing capabilities we become better integrated within the GB and so become a valuable part of it, encouraging mechanisms active in all CAS to operate in a way that prolongs our retention within this system, i.e. prolongs our biological lifespan.

**Discussion**

The fact that neurons display enhanced survival capabilities when appropriately stimulated is well accepted. For example, it has been shown that odorant stimuli selectively rescue olfactory sensory neurons from apoptosis, possibly through activation of Bcl-2 (16). Conversely, it has also been shown that when the electrical activity and excitability of neurons is reduced, these are more likely to degenerate and perish. Michel et al (17) have suggested that there is a specific need for dopaminergic neurons to be stimulated in order to survive. In addition, it is known that cognitive stimulation through an enriched environment increases adult neurogenesis and integration of neurons, as well as increase the number of neurons, their dendritic length and projections (18,19). Others have shown that any perturbation that increases the activity of the neurons will enhance the likelihood of their survival (20) and their synaptic connectivity (21). It is also known that environmental enrichment (i.e. enhanced input of information that requires action) increases several trophic factors and expression of agents such as mTOR, which enhance neuronal survival (22) and it causes positive universal brain changes (23) particularly in age-dependent degenerative conditions(24,25).

In all, it is becoming clear that the survival of the neurons is dependent upon the degree and frequency of stimulation, a fact that accords well with the observations about the survival of artificial neural networks and digital nodes. As a consequence of this, it may be suggested that the survival of noemes which are subjected to the same cybernetic laws, would also be enhanced if these are well integrated, stimulated and engaged with the whole (GB). Information-sharing can be both external (such as environmental enrichment strategies) and internal, by readjusting and reassessing internal inputs. Kalimana et al (26) have shown that meditation in humans (i.e. focused internalised information sharing), can result in changes in gene expression, and in epigenetic regulatory modulation. This was achieved through a global modification of histones, and a decreased expression of pro-inflammatory genes. This supports the view that manipulation of information (internal or external) has a direct effect on biological substrates and improves health. It was also shown that the lifespan of neurons can be independent of the survival of the host. In the absence of pathological conditions, the survival on mammalian neurons is only limited by the maximum lifespan of the organism, but when these neurones are transplanted in younger organisms, their survival continues in the longer living host(27). This may suggest that noemes could also experience increased survival which will be only limited by the survival of their host (i.e. the GB). Current opinion about the lifespan of the GB is divided, but it could be argued that this would be indefinite (but not infinite).

In order to discuss the issue further, it will be necessary to use an analogy from computer terminology. Rent's rule predicts the number of terminals acquired by a group of gates for communication with the rest of the circuit (28). In other words, it describes the relationship between the number of elements in a given area and the number of links between these elements. This can be directly applied in the case of neurons and noemes.

$T = AK^P$

T = number of links

K = number of nodes (or neurons/noemes)

A= the average number of linksin one block (brain, GB)

P=the Rent exponent

The comparison of the brain connectivity properties with computer network topologies using Rent's rule is a useful model to study hierarchical properties of both the brain (neuron) and the computer(node or gate) (29). Artificial Neural networks (ANN) are systems of interconnected nodes which can compute values from inputs and feed information through the network (clearly there is an analogy with the noeme, at a hierarchical level). The behaviour of ANN, brains and the GB is based on common principles of non-linearity, distributed, parallel processing, and adaptation. Both the brain(neuron) and the artificial neural networks (node/gate) as well as the Global Brain (noeme) are networks of 'nodes' and signal-carrying connections.The signal spreads according to the rules of Spreading Activation, i.e. if the signal is strong it survives, if it is weak it perishes. This is also true of the node. Survival of the node depends on the strength of the link and on the weight of the information (and also on the activation period, and on the number of other nodes).Spreading activation is the process whereby computer nodes (or neurons) are activated and then transmit this activation to other computer nodes (or neurons). The activation is proportional to the strength (weight) of the connections between the nodes involved. It is given by

$A(t+1) = M \times A(t)$

Where M is the matrix representing the weight of the connection between two nodes

A is the activation value of the neuron

t is the time

When the activation becomes too low the transmission of the activation may stop due to the presence of thresholds that need to be overcome. In neurons, when the activation falls below a certain threshold, the neuron is not activated and it cannot pass any signal further on. This indicates the need to maintain the activation (persistence of information over time) in order to achieve an efficient function.There is no reason to suppose that this is not also true in the case of noemes and the GB. In addition, the behaviour of a hyper-connected human agent (noeme) is subjected to the same rules as those underlying Hebbian learning (30): a persistent stimulation of one noeme by another, will result in strengthening the links between the two, inducing lasting changes which add to the agents' stability i.e. survival. In other words, there will be 'selective stabilisation' of the nodes. The links between noemes can be reinforced or inhibited, while autocatalytic processes lead to formation of novel links and strengthening of existing ones (31).

Real-world complex adaptive systems are made of networks with a scale-free degree distribution, i.e. they contain highly irregular degrees of connections. A scale-free topology is based on a power-law distribution of its elements, with many elements having a low number of connections and links, and some other elements with a very high number of connections and links. Csermely (32) suggested that highly reactive and dynamic 'creative' nodes (agents) within a network help modulate the overall functionality of the system and are thus useful to the whole. These creative agents help integrate the communication of the entire network. We can extrapolate and generalise this to include noemes as agents within a higher system, the Global Brain, which is our technological eco-system. The extrapolation suggests that noemes which help the stability and integration of the entire system (by facilitating communication between all other agents) may be retained longer, as their presence is essential to the functionality of the system. These creative elements help the entire system to invent novel solutions for continually emerging problems, integrate the network in response to unexpected environmental circumstances, and determine the system's potential for quick adaptation and evolution (32). If the system (in this case, our technological environment) experiences a high degree of fluctuations, then this will lead to an increase in the number of creative noemes. These creative elements improve the flexibility of the system and provide many degrees of freedom, enabling the system to store more useful information and thus enhance functionality, adaptation and evolvability (i.e. longer survival).

Predictions using notions from Network Science

In a well-functioning, well-connected network such as the GB, the nodes (i.e. noemes) must obey universal laws taken for Network Science, for example:

- Network diameter and Metcalf's Law. The Network Diameter is the average shortest distance between pairs of nodes. The value needs to be low, in order to allow for sufficient time for positive or negative feedback loops to operate. Network diameter = (N)/log(k), where N= number of nodes, and k=number of links (33). As the network grows, the number of links also has to grow in order to keep the network diameter low. Metcalf's Law states that the value of a telecommunications network is proportional to the square of the number of connected users of the system. Expanding Metcalf's Law to human agents we find that the value of an intelligent network is proportional to the square of the number of humans who use it, leading to a Global Brain network with a Network Diameter of 2.0. This means that humans will eventually be forming very intimate relationships with artificial intelligent networks at a rate of one to one (33).
- The Clustering Coefficient must be sufficiently large. The clustering coefficient is the ratio of existing links connecting a node's neighbours to each other, to the maximum possible number of such links. In order for this to be achieved within the Global Brain it is necessary not only for individual noemes to be well-connected, but also for the connections of those noemes to be well-connected to other noemes.
- The Connectedness: In the case of humans, an efficient network is a strongly connected network. A hypothetical collection of nodes is said to be strongly connected when there exists a directed path from any node to any other. This, applied to humans, means that ideally every human must be directly connected to all others, which is a distinct possibility, considering the predictions of Metcalf's Law mentioned above.
- Node Centrality values must be ideal. Node Centrality is a measure of 'importance' of a node within a system. It incorporates the average distance that each node is from all other nodes in the network, the number of shortest paths in a network that traverse through that node, and the number of links that a particular node possesses. Here again the important position a node occupies within a system can predict its extended survival within that system.

In practice, the above laws mean that it is necessary for humans, as the GB grows, to maintain a large number of connections, so that to maintain a low network diameter, a large numbers of links, and a high Clustering Coefficient. In order for the network to adapt and survive it must be strongly connected, with as many connections as possible, with close distances between the connections, and with a high strength of each connection.

**Conclusion**

The main thrust of this discussion is that, if computer nodes and neurons are retained longer when these are hyper-connected, then humans within a larger network must also be retained (live) longer if they are hyper-connected.Extrapolating from the above discussion, we can suggest ways to enhance hyper-connectivity of humans within the GB (Table 1).

\*\*\*\*\*\*\*\*\*\*\*\*\*\*\*\*\*\*\*\*\*\*\*\*\*\*\*\*\*\*\*\*\*\*\*\*\*\*\*\*\*\*\*\*\*\*\*\*\*\*\*\*\*\*\*\*\*\*\*\*\*\*\*\*\*\*\*\*\*\*\*\*\*\*\*\*\*\*

**TABLE 1**

**Practical possible ways to increase connections between human agents (noemes)**

- Develop a strong social media base, in diverse forums
- Stay continually visible on line
- Be respected and valued in the virtual environment
- Increase the number of your connections both in virtual and in real terms.
- Increase the unity of your connections by using only one (user)name for all environments and across all platforms.
- Increase the strength of your connections by sharing meaningful information that requires action

\*\*\*\*\*\*\*\*\*\*\*\*\*\*\*\*\*\*\*\*\*\*\*\*\*\*\*\*\*\*\*\*\*\*\*\*\*\*\*\*\*\*\*\*\*\*\*\*\*\*\*\*\*\*\*\*\*\*\*\*\*\*\*\*\*\*\*\*\*\*\*\*\*\*\*\*\*\*

Based on these common underlying principles, it is now possible to propose a novel cybernetic law, the 'Law of Requisite Usefulness'. This states that the duration of retention of an agent within a CAS is proportional to the contribution of that agent to the overall adaptability of the system.The more 'useful' an agent is within a system, with respect to the system's evolvability, adaptability and function, the more likely it is that this agent will be retained within the system, and thus survive longer.In the case of humans as agents within a global technological society, it can be concluded that the more indispensable a human is to the evolvability of the system, the longer this human will live. How this can be translated in concrete biological terms which may explain the increased physical longevity is yet unclear. Nevertheless, attempts havealready been made to elaborate on some possible biological mechanisms (5) involving information sharing processes which force a re-allocation of resources from germline cells to somatic cells. In addition, it has been shown that appropriate external stimulation may cause epigenetic re-programming, where an adult cell (ageing) can revert to a pluripotent (young) stem cell(34), and even that this epigenetic information may be transmitted to the offspring (35). This provides additional 'support in principle' to the ideas discussed above, showing that epigenetic influence (caused by continual exposure to information) may result in *de novo* youthful cells. Future research could elaborate further on other possible background biological processes that are forced to operate following a relentless information-sharing process and its impact upon human longevity.

References

1. Collier J. Entropy in evolution. Biology and Philosophy 1986;1:5-24

2. Heylighen F. The Global Superorganism: an evolutionary-cybernetic model of the emerging network society, Social Evolution & History. 2007;6 No. 1, 58-119

3. Dix, A. The brain and the web – intelligent interactions from the desktop to the world. 2006 Keynote at IHC 2006, Natal, Brazil

4. Cilliers P. Complexity and post-modernism. New York NY, Routledge 1998

5. Kyriazis M. Reversal of informational entropy and acquisition of germ-like immortality by somatic cells. Current Aging Science 2014, in press

6. Adami C, Ofria C, Collier TC. Evolution of biological complexity PNAS 2000; vol. 97(9) 4463–4468

7. Frieden BR, Gatenby RA. Information Dynamics in Living Systems: Prokaryotes, Eukaryotes, and Cancer. PLoS ONE 2011;6(7): e22085

8. Heylighen F. Principles of Systems and Cybernetics: an evolutionary perspective, in: Cybernetics and Systems, R. Trappl (ed.), World Science, Singapore,1992 p. 3-10

9. Paperin G. Green D, Sadedin S. Dual-phase evolution in complex adaptive systems. J. R. Soc. Interface 2011;8(58): 609-629

10. Pack Kaelbling L, Littman M, Moore A. Reinforcement Learning: A Survey, Journal of Artificial Intelligence Research 4, 1996;237-285

11. Teng CY, Adamic L. Longevity in Second Life. Proceedings of ICWSM, 2010

12. Brandtzaeg PB, and Heim, J. User loyalty and online communities: why members of online communities are not faithful. In INTETAIN 2008, 1-10

13. Bryant SL, Forte A, Bruckman A, Becoming wikipedian: transformation of participation in a collaborative online encyclopedia. In GROUP 2005, 1-10 ACM

14. Ramírez-Rodríguez G, Ocaña-Fernández MA, Vega-Rivera NM, Torres-Pérez OM, Gómez-Sánchez A, Estrada-Camarena E, Ortiz-López L. Environmental enrichment induces neuroplastic changes in middle age female BalbC mice and increases the hippocampal levels of BDNF, p-Akt and p-MAPK1/2. Neuroscience. 2014;260:158-70

15. Heylighen, F. Accelerating socio-technological evolution: from ephemeralization and stigmergy to the global brain. In: Globalization as evolutionary process: modelling global change. Routledge, 2008 p. 284

16. Watt WC, Sakano H, Lee ZY, et al. Odorant stimulation enhances survival of olfactory neurons via MAPK and CREB. Neuron 2004;41(6):955-967

17. Michel PP, Toulorge D, Guerreiro S, Hirsch EC. Specific needs of dopamine neurons for stimulation in order to survive: implication for Parkinson disease. FASEB J 2013;27(9):3414-23

18. Valero J, España J, Parra-Damas A, Martín E, Rodríguez-Álvarez J, Saura CA. Short-term environmental enrichment rescues adult neurogenesis and memory deficits in APP (Sw,Ind) transgenic mice. PLoS One 2011; 6(2):e16832

19. Aumann TD, Tomas D, Horne MK Environmental and behavioral modulation of the number of substantia nigra dopamine neurons in adult mice. Brain Behav. 2013;3(6):617-25

20. Lin CW, Sim S, Ainsworth A, Okada M, Kelsch W, Lois C. Genetically increased cell-intrinsic excitability enhances neuronal integration into adult brain circuits. Neuron. 2010;65(1):32-9

21. Sim S, Antolin S, Lin CW, Lin Y, Lois C. Increased cell-intrinsic excitability induces synaptic changes in new neurons in the adult dentate gyrus that require Npas4. J Neurosci. 2013;33(18):7928-40

22. Rizzi S, Bianchi P, Guidi S, Ciani E, Bartesaghi R. Impact of environmental enrichment on neurogenesis in the dentate gyrus during the early postnatal period. Brain Res. 2011;1415:23-33

23. Alwis DS, Rajan R. Environmental enrichment causes a global potentiation of neuronal responses across stimulus complexity and lamina of sensory cortex. Front Cell Neurosci. 2013 ;7:124

24. Speisman RB, Kumar A, Rani A, Pastoriza JM, Severance JE, Foster TC, Ormerod BK. Environmental enrichment restores neurogenesis and rapid acquisition in aged rats. Neurobiol Aging. 2013;34(1):263-74

25. Barone I, Novelli E, Piano I, Gargini C, Strettoi E.  Environmental enrichment extends photoreceptor survival and visual function in a mouse model of retinitis pigmentosa. PLoS One. 2012;7(11):e50726

26. Kalimana P,  Jesus Alvarez-Loopez M,  Cosın-Tomas M, Rosenkranz M, Lutzc A, Davidson R.Rapid changes in histone deacetylases and inflammatory gene expression in expert meditators. Psychoneuroendocrinology 2014;40, 96—107

27. Magrassi L, Leto K, Rossi F. Lifespan of neurons is uncoupled from organismal lifespan. Proc Natl Acad Sci U S A. 2013 12;110(11):4374-9

28. Christie P. The interpretation and application of Rent's rule. VLSI Systems 2000, IEEE Transactions 8(6) 639-648

29. Beiu V. On two-layer brain-inspired hierarchical topologies- a Rent's rule approach. Transactions on High-performance embedded architectures and compliers IV 2011;vol 6760:311-333

30. Hebb, D.O. The Organization of Behavior. New York: Wiley & Sons 1949

31. de Rosnay, J. The Symbiotic Man: A New Understanding of the Organization of Life and a Vision of the Future, McGraw-Hill Professional Publishing, 2000

32. Csermely P. Creative elements: network-based predictions of active centres in proteins and cellular and social networks. Cell 2008 doi 10.1016/j.tibs.2008.09.006 p 569-577

33. Hibbard B: Network Diameter and Emotional Values in the Global Brain. Global Brain Workshop, July 2001 Free University of Brussels

34.  Obokata H, Wakayama T, Sasai Y et al. Stimulus-triggered fate conversion of somatic cells into pluripotency. Nature 2014;505. 641-647

35. Dias BG, Ressler KJ. Parental olfactory experience influences behavior and neural structure in subsequent generations. Nat Neurosci. 2014;17(1):89-96